# Electrical Manipulation of Spin Splitting Torque in Altermagnetic $RuO_2$

Yichi Zhang[1,2], Hua Bai[1], Lei Han[1], Jiankun Dai[1], Chong Chen[1], Shixuan Liang[1], Yanzhang Cao[1], Yingying Zhang[1], Qian Wang[1], Wenxuan Zhu[1], Feng Pan[1], Cheng Song[1,*]

[1]*Key Laboratory of Advanced Materials (MOE), School of Materials Science and Engineering, Tsinghua University, Beijing 100084, China.*
[2]*Key Laboratory of Materials Modification by Laser, Ion and Electron Beams (MOE), Dalian University of Technology, Dalian 116024, China.*
[*]Corresponding author. Email: songcheng@mail.tsinghua.edu.cn (C.S.)

**Abstract**

Due to nonrelativistic altermagnetic spin splitting effect (ASSE), altermagnets can generate time-reversal-odd spin current and spin splitting torque (SST) with spin polarization parallel to the Néel vector. Hence the effective manipulation of SST would provide plenty of opportunities for designable spintronic devices, which remains elusive. Here, the electrical control of SST is achieved in altermagnetic $RuO_2$, based on controllable Néel vector of $RuO_2$ and Néel vector-dependent generation of SST. We demonstrate the current-induced switching of Néel vector via spin-orbit torque in $RuO_2$ films, according to the reversible polarity of electrical transport measurements and X-ray magnetic linear dichroism (XMLD). The XMLD also unprecedentedly demonstrates that Néel vector really exists in altermagnets. The switching of Néel vector to the current direction and resultantly enhanced spin polarization parallel to the Néel vector brings about stronger ASSE-induced spin current. Our findings not only enrich the properties of altermagnets but also pave the way for high speed memories and nano-oscillators with excellent controllability and efficiency.

**Introduction**

In the field of information storage, nonvolatile magnetic random-access memory (MRAM) with high speed, high density and low dissipation plays a key role, and it requires efficient approaches, especially electrical ones, to recording data[1–4]. During the past decades, great progress has been made in spintronics, and spin torques carrying angular momentum has replaced current-induced Oersted field as the state of the art for writing technology[5–8]. The widely used spin torques include spin transfer torque (STT) and spin-orbit torque (SOT), and corresponding STT-MRAM and SOT-MRAM possess the strengths of lower power consumption, higher storage density and better reliability[9–11]. On one hand, the STT usually has higher spin torque efficiency due to nonrelativistic ferromagnetic exchange splitting, but relevant devices are more fragile because of the writing current flowing directly through the magnetic tunneling junctions (MTJs)[12,13]. On the other hand, for SOT-MRAM, although the writing current does not pass through MTJs, the spin torque efficiency is restricted by spin-orbit coupling (SOC) of spin source materials[14,15]. Therefore, it will be much beneficial to the improvement of data writing technique if the advantages of both STT and SOT can be combined together.

Recently, a new spin splitting torque (SST) related with nonrelativistic altermagnetic spin splitting effect (ASSE) has been predicted theoretically in collinear antiferromagnets with crystal and spin-rotation symmetries (Supplementary Materials Note S1), which are termed as altermagnets[16–21]. This SOC-independent ASSE is able to generate nontrivial transverse spin current or out-of-plane polarized spin current with higher efficiency compared to other mechanisms, such as low crystal symmetry[22–24]. Experimental evidence of SST has been reported in $RuO_2$ films with rutile crystal structure[25–27]. The SST exhibits controllable spin polarization with its direction parallel to the Néel vector of the altermagnets, which can overcome the limitation of orthogonal relation among charge current, spin polarization and spin current. Not only does $RuO_2$ possess desirable electrical conductivity despite being a metallic oxide, but it also has relatively high Néel temperature above 300 K as well as feasible preparation techniques, which are in favor of being applied in spintronic devices operating at room temperature[28–32].

Most notably, the Néel vector of $RuO_2$ aligns along the [001] crystallographic axis, and hence various behaviors related with SST can be observed in $RuO_2$ films with different crystallographic orientations[28,29]. For example, $RuO_2$(101) films with tilted Néel vector are not only able to generate SST with out-of-plane polarization in favor of field-free switching of the magnetization in adjacent magnetic layer[26,27], but they can also be utilized for spin detection[33]. To extend the functionalities of altermagnetic spin sources, SST is expected to be manipulated, which will be achieved once the Néel vector can be controlled efficiently in altermagnets. Meanwhile, the effective control of Néel vector is likely to change the band structure of altermagnets, which will also enrich the new physics of altermagnetic materials[34]. Taking into account the difficulty of using extremely large magnetic field (usually dozens of Tesla) to overcome the anisotropic field of $RuO_2$, it is more applicable to adopt electrical approach as an alternative solution, which is also suitable for device integration[35,36]. In this work, we achieve the effective manipulation of Néel vector to tune SST in altermagnetic $RuO_2$ via electrical approach. More importantly, both (i) controllable Néel vector under electrical currents and (ii) Néel vector-dependent generation of spin current and SST are two crucial prerequisites to accomplish this task. As presented in Fig. 1a, a $y$-direction preset current ($J_{preset}$) flowing inside heavy metal (HM) layer generates SOT and switches the Néel vector ($N$) in $RuO_2$ layer to be along $y$-axis as well. The charge current ($J_C$) in $RuO_2$ layer produces SST with spin polarization parallel to $N$, which is along $y$-axis ($\sigma_y$). By comparison, if using another $J_{preset}$ along $x$-axis at first, we can align $N$ along $x$-axis and then obtain adjustable SST with $x$-direction spin polarization ($\sigma_x$) by applying $J_C$, as displayed in Fig. 1b.

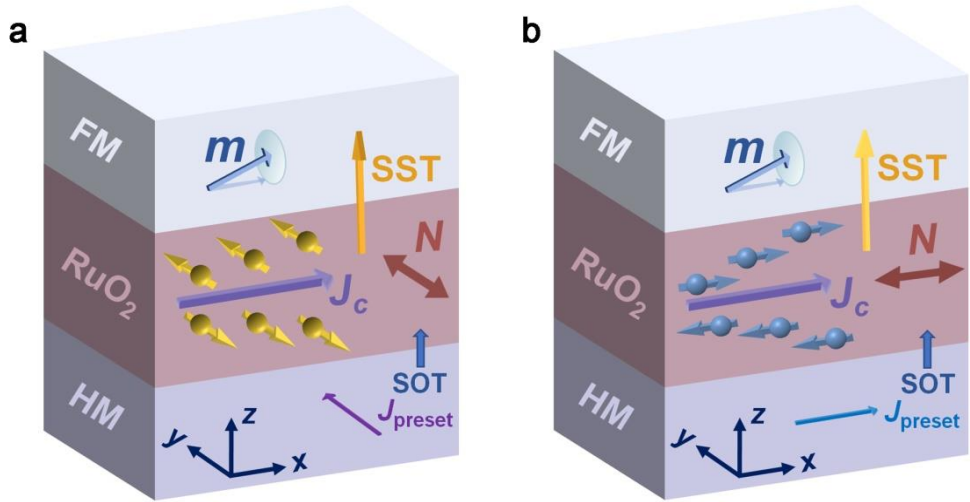

**Fig. 1. Electrical manipulation of spin splitting torque in altermagnetic RuO$_2$ with two operation schemes**. **a** The *y*-direction $J_{preset}$ flowing in subjacent HM layer generates SOT, leading to the alignment of *N* to be parallel with *y*-axis in RuO$_2$ layer, and the charge current in RuO$_2$ layer produces SST with $\sigma_y$. **b** The *x*-direction $J_{preset}$ switches *N* to be along the *x*-direction, giving rise to tunable SST with $\sigma_x$.

## Results

### Growth and fundamental characterizations of film samples

According to the physical picture of ASSE (Supplementary Materials Fig. S1), RuO$_2$(100) can generate SST by applying charge current while RuO$_2$(110) cannot, though the Néel vector lies in plane for both cases. To investigate SST manipulation in altermagnetic RuO$_2$, we need to achieve controllable Néel vector in RuO$_2$ films with in-plane easy axis. Hence we deposited Pt/RuO$_2$ film samples on Al$_2$O$_3$(0001) and MgO(100) substrates, respectively. The X-ray diffraction (XRD) patterns indicate that RuO$_2$ has strong (100) and (110) texture in Al$_2$O$_3$/Pt/RuO$_2$ and MgO/Pt/RuO$_2$ samples respectively, and they exhibit fine crystallization (Supplementary Materials Fig. S2). The straight lines of magnetic field-dependent magnetization (*M-H*) curves measured by superconducting quantum interference device (SQUID) at 300 K (Supplementary Materials Fig. S3) can be accounted for by the diamagnetism of substrates, and there is negligible net magnetization in Pt/RuO$_2$(100) or Pt/RuO$_2$(110). Besides, the temperature-dependent magnetization (*M-T*) curves measured by SQUID reveal the Néel temperature of $RuO_2$ films on $Al_2O_3$ and MgO substrates to be 400~450 K (Supplementary Materials Fig. S4). The $Al_2O_3$/Pt/$RuO_2$ and MgO/Pt/$RuO_2$ film samples with desirable crystal quality and magnetization can be used for further depositing comparable Py layers, which are also the basis for conducting following experiments on various device samples (Supplementary Materials Fig. S5).

**Current-induced Néel vector switching of $RuO_2$**

First we investigate current-induced switching of Néel vector in altermagnetic $RuO_2$ by electrical transport measurement at room temperature. We fabricated eight-terminal devices of Pt(5 nm)/$RuO_2$(15 nm) on $Al_2O_3$ and MgO substrates. As shown in the inset of Fig. 2a, each device has a pair of vertical ($I_{write1}$) and horizontal ($I_{write2}$) channels with the width of 20 μm. After applying each writing current pulse with different magnitude along one certain channel, the change of transverse Hall resistance ($\Delta R_{Hall}$) is plotted in Fig. 2a for $RuO_2$(100) (red square) and $RuO_2$(110) (blue diamond). When we increase the magnitude of applied current pulses, no remarkable $\Delta R_{Hall}$ signals can be observed until it reaches threshold value, corresponding with the current densities of $5.3\times10^{11}$ A $m^{-2}$ and $4.8\times10^{11}$ A $m^{-2}$ in Pt layer ($J_{Pt}$), for Pt/$RuO_2$(100) and Pt/$RuO_2$(110) samples respectively[27,37]. These critical values can be taken as reference to conduct the following experiments. The slight discrepancy of critical switching current densities is mainly caused by the different anisotropy when the films were deposited on different substrates. Then we use critical switching currents to test each Pt/$RuO_2$ device for three cycles: a group of five successive pulse currents with the same amplitude are applied alternatively along the two orthogonal channels. Figure 2b displays $\Delta R_{Hall}$ of $RuO_2$(100) (red square) and $RuO_2$(110) (blue diamond) as a function of pulse numbers. For the two Pt/$RuO_2$ devices, the obvious $\Delta R_{Hall}$ is observed after applying the first pulse current, and it increases gradually regarding the second to the fifth pulses. By comparison, no apparent $\Delta R_{Hall}$ signals can be detected in Pt or $RuO_2$ devices (Supplementary Materials Note S4). Significantly, previous literature has reported that thermal artifacts could be observed in electrical Hall resistance measurements under specific circumstances[38].

Hence to study the Joule heating caused by applied current pulses[39], we meticulously conducted both experiment measurement and finite element simulation, and the results distinctly demonstrate that the temperature of device samples remains under the Néel temperature from start to finish (Supplementary Materials Note S5). Therefore, the recorded $\Delta R_{Hall}$ of Pt/RuO$_2$ devices can be attributed to current-induced Néel vector switching. The Hall resistance measurements preliminarily prove the in-plane Néel vector switching with 90° in altermagnetic RuO$_2$, where partial switching of domains is likely to exist here[40–43].

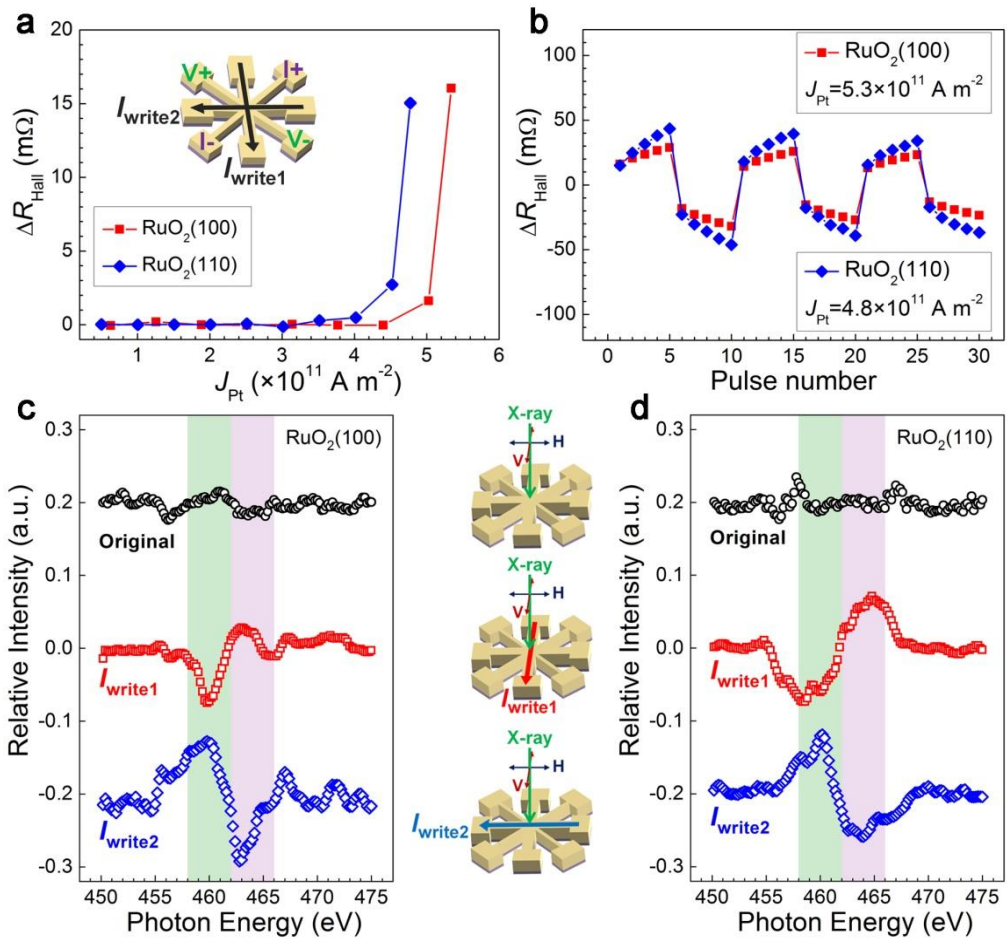

**Fig. 2. Current-induced Néel vector switching in altermagnetic RuO$_2$ probed by electrical transport and XMLD measurements. a** $\Delta R_{Hall}$ as a function of current density in the Pt layer, with each data point being collected after applying one pulse current. The inset shows the schematic of eight-terminal device for electric transport measurement, where the read current and voltage paths are mutually perpendicular and are aligned 45° to the two writing channels. **b** Pulse number dependence of $\Delta R_{Hall}$ of Pt/RuO$_2$(100) (red square) and Pt/RuO$_2$(110) (blue diamond) devices under critical switching current densities. Ru $M$-edge XMLD results of Pt/RuO$_2$ devices for **c** RuO$_2$(100) and **d** RuO$_2$(110). From top to bottom: before applying writing currents (Original), after applying a group of five writing currents along vertical ($I_{write1}$) channel and along horizontal ($I_{write2}$) channel.

To further verify the in-plane Néel vector switching in altermagnetic RuO$_2$ films, we performed the measurement of Ru $M$-edge X-ray magnetic linear dichroism (XMLD) at room temperature. By this synchrotron radiation test technique which is sensitive to in-plane Néel vector, we can reveal the distribution of Néel vector in RuO$_2$(100) and RuO$_2$(110) and its dependence on writing currents[44–46]. Considering the compatibility with the size of detecting spot (80×80 μm$^2$) for XMLD measurement, we fabricated the eight-terminal devices with 100 μm-wide writing channels here for Pt(5 nm)/RuO$_2$(15 nm) in which the crystalline orientations of RuO$_2$ are (100) and (110). The detecting light is focused on the intersection area of the device, and the signals of linearly polarized X-ray absorption spectroscopy (XAS) are collected (Supplementary Materials Fig. S11 and Fig. S12). The peak position of XAS curves near 462 eV corresponds to the $M_3$ $3p_{3/2}$ binding energy of Ru element. The XMLD results presented in Fig. 2c and Fig. 2d are obtained by the difference of linearly vertical (V) and horizontal (H) polarized XAS signals.

Before applying writing currents, we collected XMLD signals of the initial state of both devices. Due to non-epitaxial growth mode, twin crystal structure with different in-plane alignment of Néel vector exists in RuO$_2$ films[25]. Hence the in-plane Néel vectors without uniaxial orientation cancel out each other, leading to no peaks or valleys in XMLD results for the initial state of both switching devices, which is presented by the (black circle) curves at the top of Fig. 2c and Fig. 2d. For each eight-terminal device, writing currents are applied by following two schemes: a group of successive five pulse currents flows through either the vertical channel ($I_{write1}$) or the horizontal channel ($I_{write2}$). The values of $J_{Pt}$ are 5.4×10$^{11}$ A m$^{-2}$ and 4.9×10$^{11}$ A m$^{-2}$ for Pt/RuO$_2$(100) and Pt/RuO$_2$(110) devices, respectively. To ensure the repeatability of testing results, we have measured sufficient switching devices after applying writing currents. Noticeably, spikes appear in the XMLD signals after applying writing currents, as evidenced by the representative peaks and valleys in middle (red square) and bottom (blue diamond) curves in Fig. 2c and Fig. 2d. After $I_{write1}$ flows through the vertical channel of Pt/RuO$_2$(100) and Pt/RuO$_2$(110) devices, the XMLD curves undergo zero-negative-positive-zero trace as photon energy increases, as displayed by two (red square) curves in the middle of Fig. 2c and Fig. 2d. Meanwhile, after we applied $I_{write2}$ to these devices, the distinct XMLD signals can be seen in the two (blue diamond) curves at the bottom of Fig. 2c and Fig. 2d. It is reasonable to deduce that after we used writing currents which reached the threshold, the Néel vector is capable of being switched towards the direction of charge current in Pt/RuO$_2$ devices. In contrast to their counterparts in the middle, the two curves at the bottom of Fig. 2c and Fig. 2d reveal that the XMLD signals exhibit a zero-positive-negative-zero tendency with increasing photon energy, which indicates the reversed polarity of XMLD signals after we rotated the direction of writing currents with in-plane 90°. Of note that there is a delay time from the application of current pulses and the measurement of XAS (or XMLD) signals. We have prolonged the delay time to make another test and observed the same phenomenon in XMLD patterns, which proves the stability of XMLD measurement results (Supplementary Materials Fig. S13). Therefore, the onefold orientation of Néel vector after applying writing currents is determined by the direction of applied charge currents, which results from SOT-induced Néel vector switching. The above XMLD measurements of Ru $M$-edge support the current-induced switching of Néel vector with in-plane 90° in altermagnetic RuO$_2$, which is an important basis for manipulating SST because of the close relation between SST and the state of Néel vector. The XMLD measurements also unprecedentedly demonstrate that Néel vector really exists in the altermagnets, besides the previous evidence deduced by transport measurements, such as the anomalous Hall effect and SST[25–27,35].

**Manipulation of spin splitting torque in RuO$_2$**

To study the electrical control of spin splitting torque related to spin current generation in altermagnetic RuO$_2$, we fabricated Pt/RuO$_2$(100)/Py and Pt/RuO$_2$(110)/Py devices for spin torque ferromagnetic resonance (ST-FMR) measurements, where the thickness of each layer is indicated by Pt(5 nm)/RuO$_2$(10 nm)/Py(15 nm). It is worth noting that this well-established ST-FMR technique is particularly appropriate for detecting and analyzing spin currents with various spin polarizations ($\sigma_x$, $\sigma_y$ and $\sigma_z$)[23,26,47,48]. Figure 3a illustrates the setup of ST-FMR measurement to characterize the SST in Pt(5 nm)/RuO$_2$(10 nm)/Py(15 nm) devices at room temperature. A radio-frequency (RF) charge current flowing through the RuO$_2$ layer is converted into transverse spin current, and then it is injected into the adjacent Py layer. According to the physical picture of ASSE (Supplementary Materials Fig. S1), the spin-current generation is contributed by ASSE together with spin Hall effect (SHE) in Pt/RuO$_2$(100)/Py, while it is due solely to SHE in Pt/RuO$_2$(110)/Py. Under the applied magnetic field, the oscillating spin current exerts spin torques on ferromagnetic Py layer leading to magnetization precession. The external magnetic field is swept with $\varphi_H$ (angle between charge current and applied magnetic field), which can be regarded the same as $\varphi_M$ (angle between charge current and magnetization in Py) because of the relatively small in-plane anisotropy of Py. Figure 3b displays a typical ST-FMR spectrum of Pt/RuO$_2$(100)/Py device after applying a group of writing currents ($I_\text{write}$) of 60 mA, and it is measured with $\varphi_H = 45°$ under the excitation of 6 GHz and 19 dBm microwave current. As shown in Fig. 3b (with more same type of data in Supplementary Materials Fig. S14), the (black) dots are measured DC voltage data, and corresponding fitted $V_\text{mix}$ curve (green one) can be decomposed into both symmetric ($V_\text{S}$, blue one) and antisymmetric ($V_\text{A}$, red one) Lorentz line shapes. From the voltage signals, we can see that the amplitude of $V_\text{A}$ (red line) is comparatively much larger than that of $V_\text{S}$ (blue line), for the Oersted field plays a remarkable role in the former resulting from the shunting effect.

Each time after applying a group of five successive writing currents with the same magnitude to Pt/RuO$_2$/Py devices, we conducted an angle-dependent ST-FMR measurement to characterize the SST in altermagnetic RuO$_2$. The magnetic field was applied in the plane of tested samples, and the angle between the Pt/RuO$_2$/Py devices and magnetic field was defined as $\varphi_H$. We swept $\varphi_H$ with a step size of 15° under 6 GHz and 19 dBm microwave currents. Of note is that RuO$_2$(100) and RuO$_2$(110) studied in the present case can only generate in-plane spin polarizations ($\sigma_x$ and $\sigma_y$). Besides, $V_S$ and $V_A$ are related with in-plane and out-of-plane current-induced spin torques respectively (Supplementary Materials Table S1). Therefore, we are able to analyze spin torques with in-plane polarization directions in Pt/RuO$_2$(100)/Py device by trigonometric function fitting, according to the acquired $V_S$ data. The fitting results after applying $I_{write}$ = 60 mA and $I_{write}$ = 150 mA are presented in Fig 3c and Fig 3d, which are below and above the critical switching current density, respectively. As depicted by (purple and orange) curves in Fig 3c and Fig 3d, prominent $S_{DL}^{X}$ can be observed in the latter, while $S_{DL}^{Y}$ remains almost the same under the two circumstances.

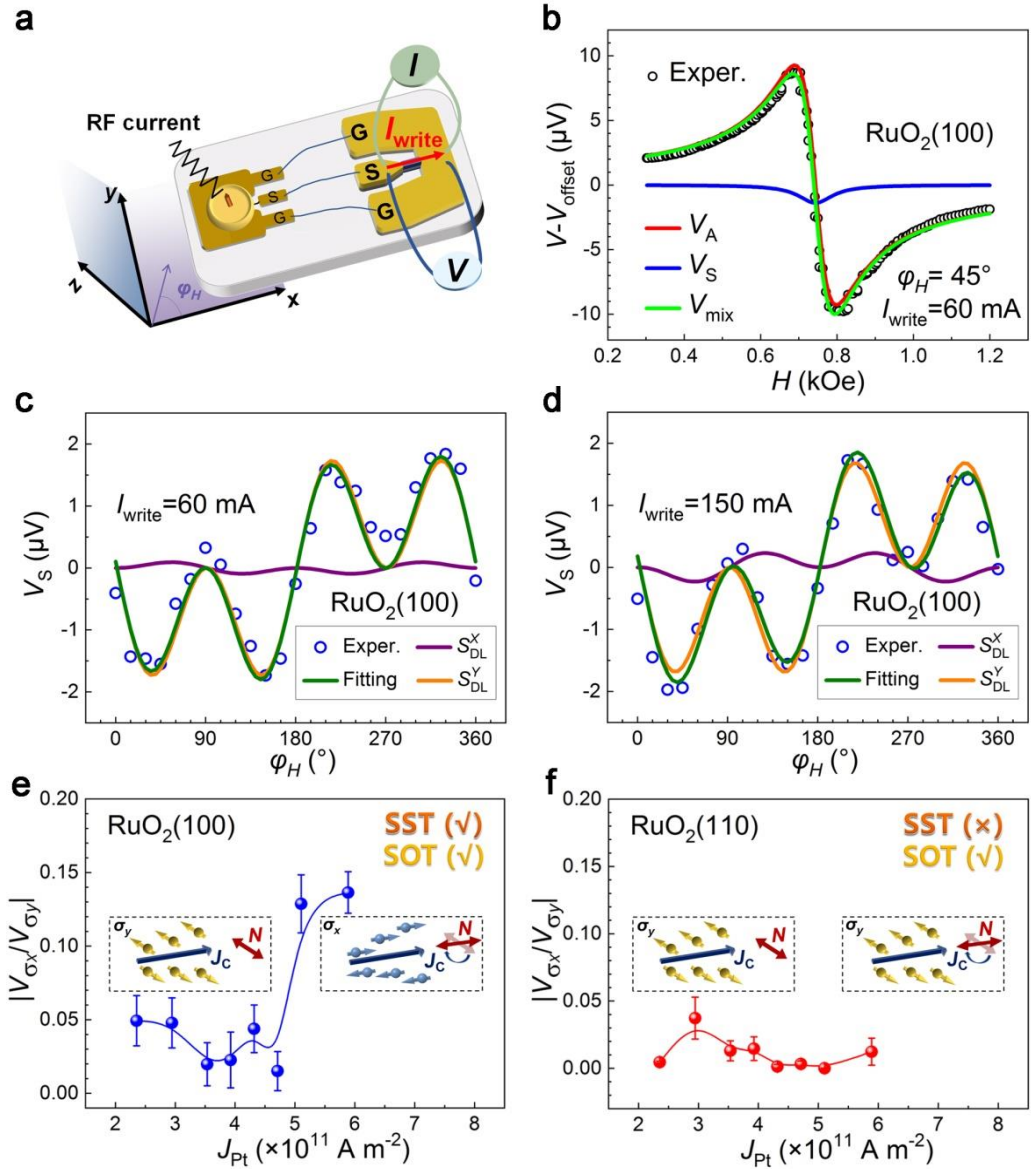

**Fig. 3. Manipulation of spin splitting torque in altermagnetic $RuO_2$ characterized by ST-FMR. a** The measurement setup of ST-FMR. **b** A typical ST-FMR spectrum consisted of antisymmetric ($V_A$, red curve) and symmetric ($V_S$, blue curve) signals under $\varphi_H = 45°$ after applying writing currents of 60 mA. The analysis of spin torques with various polarization directions by fitting the angle-dependent $V_S$ results after applying writing currents of **c** 60 mA and **d** 150 mA. All data in panels **b**, **c** and **d** are obtained by measuring Pt/$RuO_2$(100)/Py device under 6 GHz and 19 dBm. The $|V_{\sigma x}/V_{\sigma y}|$ data points of **e** Pt/$RuO_2$(100)/Py and **f** Pt/$RuO_2$(110)/Py devices are displayed in terms of different applied writing currents. The $|V_{\sigma x}/V_{\sigma y}|$ indicates the ratio of dampinglike torques with x-direction polarization ($\sigma_x$) to those with y-direction polarization ($\sigma_y$). The

data points with error bars in panels **e** and **f** are fitted by B-spline curves, and the insets refer to the state of SST before and after Néel vector switching in Pt/RuO$_2$(100)/Py and Pt/RuO$_2$(110)/Py devices.

Notably, both SST (by ASSE) and SOT (by SHE) give rise to $\sigma_y$, but $\sigma_x$ can only be attributed to SST. Therefore, we extract the ratios of $S_{DL}^X$ to $S_{DL}^Y$ from the angle-dependent measurements as $|V_{\sigma x}/V_{\sigma y}|$ in terms of the writing currents. Correspondingly, the data points of $|V_{\sigma x}/V_{\sigma y}|$ connected by B-spline lines are presented in Fig. 3e and Fig. 3f for Pt/RuO$_2$(100)/Py and Pt/RuO$_2$(110)/Py, respectively. Taking into account the multiple steps of data fitting with possible propagation of fitting errors, error bars are also displayed accompanying the final data points in Fig. 3e and Fig. 3f. As the magnitude of writing currents rises, the $|V_{\sigma x}/V_{\sigma y}|$ for Pt/RuO$_2$(100)/Py fluctuates within a reasonable range, and then it increases abruptly at $I_{\text{write}}$ = 130 mA, as displayed in Fig. 3e. Under this situation, the corresponding writing current density is about 5.1×10$^{11}$ A m$^{-2}$, which reaches the threshold writing current density to trigger Néel vector switching. This result is further verified by repeated experiment (Supplementary Materials Fig. S15), whose critical value of current densities is also in accordance with the previous results measured by SOT switching experiments. Hence this current-induced Néel vector switching in RuO$_2$ layer leads to the alignment of Néel vector to be along *x*-axis (being parallel to the direction of writing currents). Considering the potential partial switching of Néel vector, further improvement of switching ratio of domains will probably enhance the value of $|V_{\sigma x}/V_{\sigma y}|$ to a larger extent. By contrast, Fig. 3f shows that there is no apparent enhancement for the $\sigma_x$ component in Pt/RuO$_2$(110)/Py as the writing currents increase, and typical angular-dependent ST-FMR results of Pt/RuO$_2$(110)/Py (Supplementary Materials Fig. S16) are also different from those of Pt/RuO$_2$(100)/Py shown in Fig 3c and Fig 3d.

**Discussion**

In view of the situation where the Néel vector lies in plane, RuO$_2$(100) films have higher spin torque efficiency than RuO$_2$(110) films do because of the existence of SST in the former, and the resultant spin torque conductivity surpasses that of many heavy metals and topological insulators, which are potential candidates for efficient spin sources[10,14,25,49]. However, the spin polarization directions of SST are unable to be changed in previous studies. To address this issue, a feasible solution is to manipulate the Néel vector in altermagnets by electrical currents. Once the writing currents meet the threshold, the SST can be tuned by applying charge currents in altermagnetic RuO$_2$(100) layer along *x*-axis: its spin polarization direction is parallel to the switched Néel vector, which is shifted from $\sigma_y$ to $\sigma_x$ (Fig. 3e). After the Néel vector is switched in RuO$_2$(100), the proportion of $\sigma_x$ enhances but $\sigma_y$ still dominates, which can be ascribed to conventional spin Hall effect. By comparison, no SST can be generated in (110)-oriented RuO$_2$ by applied charge currents, according to the physical picture of ASSE mechanism (Fig. 3f). Hence the generated transverse spin current comes entirely from the conventional SHE depending on spin-orbit coupling, and the spin torques regarding different in-plane coordinate axes cannot be tuned, which is irrelevant to the Néel vector switching in RuO$_2$(110).

It has been studied that magnetic moment-dependent spin torques can be produced by magnetic SHE (e.g., in Mn$_3$Sn) or antiferromagnetic SHE (e.g., in Mn$_2$Au), abbreviated as MSHE or AFM-SHE respectively[24,50]. Therefore, altermagnetic materials including RuO$_2$ with ASSE offers a new way to generate the spin torques besides the above two counterparts. Meanwhile, the electric field control of Néel vector in collinear antiferromagnet Mn$_2$Au with AFM-SHE is also able to tune the Néel spin-orbit torque (NSOT), which is different from the case in non-collinear Mn$_3$Sn[50,51]. Sharing the similarity of Néel vector-dependent generation of spin torques, the tunable spin polarization of SST is parallel with the Néel vector in altermagnetic RuO$_2$ while that of NSOT is perpendicular to the Néel vector in Mn$_2$Au. Furthermore, compared to Mn$_2$Au relying on multi-step Rashba-like mechanism, our RuO$_2$ depends on nonrelativistic altermagnetic spin splitting to generate spin current and spin splitting torque with tunable polarization and higher efficiency.

In summary, we successfully achieve the electrical manipulation of altermagnetic

RuO$_2$ to tune the spin splitting torque, which is accomplished based on (i) controllable Néel vector of RuO$_2$ and (ii) Néel vector-dependent generation of spin current and SST due to ASSE. By applying electrical writing currents, the Néel vector in altermagnetic RuO$_2$ with in-plane easy axis is switched by 90°. This current-induced switching is evidenced by electrical transport measurements of transverse Hall resistance and it is also manifested unambiguously by XMLD results in terms of the writing currents (and their directions). The XMLD results also unprecedentedly demonstrate that Néel vector really exists in altermagnets. Moreover, the spin polarizations of SST along different coordinate axes are analyzed by the angular dependence of ST-FMR voltage signals, and it proves distinctly that the SST can be tuned based on the control of Néel vector in altermagnetic RuO$_2$. The effective manipulation of spin splitting torque in altermagnetic RuO$_2$ via a feasible electrical approach not only enriches the knowledge of manipulating altermagnets with altermagnetic spin splitting effect, but it is also beneficial to expanding the potential applications of altermagnetic spin sources with controllable SST as the efficient writing technology in the field of information storage involving MRAM.

## Methods

### Sample preparation

By DC magnetron sputtering method, two groups of Pt(5 nm)/RuO$_2$(15 nm) film samples were prepared on single-crystal Al$_2$O$_3$(0001) and MgO(100) substrates, and single Pt(5 nm) and RuO$_2$(15 nm) films were also deposited directly on the two substrates. The Pt layers were deposited by sputtering Pt target with Ar flow of 20 at 773 K under a base pressure below $5\times10^{-5}$ Pa, and then they were annealed at 973 K for 1 hour. The RuO$_2$ layers were in-situ deposited by sputtering Ru target with Ar:O$_2$ flow of 20:5 at 773 K. The deposition rates of Pt and RuO$_2$ films were 0.69 nm/min and 3.3 nm/min, respectively. For ST-FMR measurement, we deposited Pt(5 nm)/RuO$_2$(10 nm)/Py(15 nm) film samples on Al$_2$O$_3$(0001) and MgO(100) substrates. After depositing Pt and RuO$_2$ layers with the parameters introduced above, we in-situ sputtered Py target with Ar flow of 20 at room temperature with the deposition rate of 1.33 nm/min. Also, the Pt(5 nm)/RuO$_2$(10 nm)/Py(15 nm) film samples were in-situ covered by 1 nm-thick Ru to prevent them from oxidation.

**Sample characterization**

Crystal quality and phase composition were analyzed by X-ray diffraction (XRD) characterization utilizing a Rigaku Smartlab instrument with Cu-Kα radiation (wavelength = 0.154 nm). The working voltage and working current were 40 kV and 150 mA, respectively. The scanning rate was 10°/min for each XRD pattern. Magnetic field dependence of magnetization (*M-H*) curves were measured via a superconducting quantum interference device (SQUID) from Quantum Design at room temperature. Temperature dependence of magnetization (*M-T*) curves were measured via the same SQUID instrument from 300 K to 600 K. Photographs of three kinds of devices were taken by a Nikon ECLIPSE LV150NL optical microscope.

**Fabrication of devices**

By standard photolithography and Ar-ion milling techniques, the deposited Pt(5 nm), RuO$_2$(15 nm) and Pt(5 nm)/RuO$_2$(15 nm) films on single-crystal Al$_2$O$_3$(0001) and MgO(100) substrates were patterned into eight-terminal devices. The width of two orthogonal writing channels was 20 μm (for electrical transport measurement) or 100 μm (for XAS and XMLD measurement), and the two diagonal detecting paths were 5 μm wide. By standard photolithography and Ar-ion milling techniques, the deposited Pt(5 nm)/RuO$_2$(10 nm)/Py(15 nm)/Ru(1 nm) films on Al$_2$O$_3$(0001) and MgO(100) substrates were patterned into stripes whose length and width were 50 μm and 20 μm, respectively. Then the Cr(10 nm)/Cu(80 nm) electrode was deposited on the stripes via electron-beam evaporation method.

**Electrical transport measurement**

The two diagonal detecting paths of Pt(5 nm), RuO$_2$(15 nm) and Pt(5 nm)/RuO$_2$(15 nm) devices were connected to a Keitheley 2400 current source offering a reading current of 1 mA and a Keitheley 2182 nanovoltmeter recording transverse Hall voltage signals.

A Keysight 2901 power source was adopted to output current pulses, which was also able to monitor the real-time resistance change of tested devices. A group of five successive pulse currents with the same amplitude were applied along the horizontal channel (red arrow $I_{write1}$ in the inset of Fig. 2a), and then another group of writing currents were applied along the vertical channel (blue arrow $I_{write2}$ in the inset of Fig. 2a), which constituted a testing cycle. After applying each current pulse, the calculated gap value of Hall resistance with respect to the initial value before applying this group of current pulses is defined as the $\Delta R_{Hall}$

## XAS and XMLD measurements

XAS and XMLD measurements in total-electron-yield mode were carried out at the beam line BL08U1A in the Shanghai Synchrotron Radiation Facility at 300 K. We used this technique to probe into the in-plane distribution of Néel vector in altermagnetic $RuO_2$. The XMLD curves were the differences of linearly vertical (V) and horizontal (H) polarized XAS signals. The detecting light was perpendicular to the sample plane, and the spot with the size of 80×80 μm² was focused on the intersection of eight-terminal devices.

## ST-FMR measurement

We adopted ST-FMR technique to characterize generated spin current and spin splitting torque with different polarizations in altermagnetic $RuO_2$. The sample holder was connected to a Ceyear 1465F microwave source (0.1~40 GHz) via a bias-T which a Keitheley 2182 nanovoltmeter was also linked to. External magnetic field was swept along $\varphi_H$ (0° ≤ $\varphi_H$ ≤ 360°) with step size of 15°, and microwave current with 19 dBm and 6 GHz was flowing through the stripe. Meanwhile, the direct current (DC) voltage signals resulting from the anisotropic magnetoresistance (AMR) rectification of Py were recorded by the nanovoltmeter, which were consisted of symmetric and antisymmetric Lorentz line shapes, with the expression $V = V_S L_S + V_A L_A$. As for $L_A = \Delta H(H-H_{res})/[(H-H_{res})^2+\Delta H^2]$ and $L_S = (\Delta H)^2/[(H-H_{res})^2+\Delta H^2]$, $\Delta H$, $H_{res}$ and $H$ represent line width, resonance field and external magnetic field, respectively. For the spin torques due to in-plane spin polarizations, the angular dependence of $V_S$ and $V_S$ could be described by the following equations[23,26]:

$$V_S = S_{DL}^X \sin 2\varphi \sin\varphi + S_{DL}^Y \sin 2\varphi \cos\varphi \quad (1)$$

$$V_A = A_{FL}^X \sin 2\varphi \sin\varphi + A_{FL}^Y \sin 2\varphi \cos\varphi \quad (2)$$

where $S_{DL}^X$ and $S_{DL}^Y$ were the coefficients of dampinglike (DL) torques, and $A_{FL}^X$ and $A_{FL}^Y$ were those of fieldlike (FL) torques, with $A_{FL}^Y$ also including the contribution by Oersted field torque.

**Finite element simulations**

We utilized finite element simulations for analyzing the temperature rise due to applied current pulses in Pt/RuO$_2$ devices. The simulated structures were Sub./Pt(5 nm)/RuO$_2$(15 nm) where Sub. referred to single-crystal Al$_2$O$_3$(0001) and MgO(100) substrates. Meanwhile, both writing channels of the simulated devices were 20 μm wide, which was in line with the device structure in electrical transport measurement. For the two Pt/RuO$_2$ devices, we applied a group of five 1 ms-wide current pulses with critical value, and the interval between two successive pulses was 10 s. The thermal parameters of RuO$_2$, including thermal conductivities (0.50 W cm$^{-1}$ K$^{-1}$) and specific heat capacities (56.42 J mol$^{-1}$ K$^{-1}$), were obtained from literature[52,53], and those of Pt, Al$_2$O$_3$ and MgO were acquired directly from the build-in database of simulation software. The simulated results (Supplementary Materials Fig. S10) could be classified as two categories: (i) time-dependent temperature variation, and (ii) space-dependent temperature distribution. For category (i), the monitoring space point was located at the center of the Pt/RuO$_2$ switching devices (distance = 0 μm). For category (ii), the monitoring time point was in the middle of one current pulse ($t$ = 0.5 ms).

**Acknowledgments**

Professor Wanjun Jiang and Le Zhao from Tsinghua University are greatly thanked for the help with experiment. This work is supported by the National Key Research and Development Program of China (Grant No. 2021YFB3601301) and the National Natural Science Foundation of China (Grant No. 52225106, 12241404, T2394471, and 523B1007), and the Open Fund of the State Key Laboratory of Spintronics Devices and Technologies (Grants No. SPL-2401). We acknowledge the beam line BL08U1A in Shanghai Synchrotron Radiation Facility (SSRF) for the XAS and XMLD measurements and the nanofabrication by Ultraviolet Maskless lithography machine (model: Uv litho- AcA, tuotuo technology).

**Author contributions**

F.P and C.S. supervised the project. Y.Z. conceived the idea and designed the experiments. Y.Z. conducted experiments, and H.B., L.H., J.D., C.C., S.L., Y.C., Y.Z., Q.W. and W.Z. helped with the experiments. Y.Z. wrote the original draft of the paper, and H.B., L.H, and C.S. revised and edited the paper.

**Competing interests**

Authors declare that they have no competing interests.

**Data availability**

The data supporting the findings of this study are included in the paper and its Supplementary Materials file. Further data sets are available from the corresponding author on reasonable request.